\documentclass{aa}  
\usepackage{graphicx}
\usepackage{txfonts}

\usepackage{natbib}  
\bibpunct{(}{)}{;}{a}{}{,}
\usepackage{txfonts}
\usepackage[]{hyperref}
\usepackage{amsmath}
\usepackage{amstext}
\hypersetup{
  colorlinks   = true, 
 urlcolor     = blue, 
 linkcolor    = blue, 
 citecolor   = blue 
}

\newcommand{\dw}{DE1253$-$57\xspace}
\newcommand{\dww}{DE1756$-$45\xspace}
\newcommand{\dwtwo}{DE0630$-$18\xspace}
\newcommand{\dwnine}{DE0823$-$49\xspace}
\newcommand{\dwfifteen}{DE1520$-$44\xspace}

\newcommand{\palta}{PALTA-1\xspace}
\newcommand{\paltab}{\palta~b\xspace}
\newcommand{\gaia}{\emph{Gaia}\xspace}
\newcommand\gdr[1]{\gaia~DR#1}

\begin{document}
 \title{Astrometric planet search around southern ultracool dwarfs}
  \subtitle{V. Discovery of the exoplanet candidate \paltab}
\author{J.\ Sahlmann\inst{1}\corrauth{Johannes.Sahlmann@esa.int} 
\and 
P.\ F.\ Lazorenko\inst{2} 
             	 \and E. L. Mart\'in\inst{3}
                 \and D. S\'egransan\inst{4}
                 \and M. Mayor\inst{4}
             	 \and S. Udry\inst{4}
}
\institute{
European Space Agency (ESA), European Space Astronomy Centre (ESAC), Camino Bajo del Castillo s/n, 28692, Villanueva de la Ca\~nada, Madrid, Spain 
                		\and
Main Astronomical Observatory, National Academy of Sciences of the Ukraine, Zabolotnogo 27, 03680 Kyiv, Ukraine              
\and
   Instituto de Astrof\'isica de Canarias, Calle V\'ia L\'actea, 38205 San Crist\'obal de La Laguna, Spain  
     \and
    D\'epartement d'Astronomie, Universit\'e de Gen\`eve, Chemin Pegasi 51, 1290 Versoix, Switzerland
}
   \date{Received 28 March 2026; Accepted 28 September 2026}
  \abstract
  {We report the discovery of \paltab, a sub-Jupiter mass exoplanet candidate orbiting the nearby early-L dwarf DENIS-P J1253108$-$570924 (\dw) located at a distance of $\sim$17 pc. The exoplanet was discovered by monitoring the L dwarf's astrometric motion with {FORS2 at the Very Large Telescope} over more than eight years and characterising the deviations from the expected single-source motion. The estimated planet mass is between 0.54\,$M_{\rm Jup}$ and {0.77}\,$M_{\rm Jup}$,  depending on the uncertain mass of the ultracool dwarf host. The system's mass ratio is about 0.010, and the relative separation is estimated at 1.6--3.4~au. {If confirmed, \paltab would be one of the very few known examples of a giant planet orbiting a nearby ultracool dwarf.} This discovery finalises a larger astrometric planet search survey that targeted 20 nearby ultracool dwarfs and previously discovered three new ultracool binary systems.}

\titlerunning{Discovery of \paltab}
\maketitle
\nolinenumbers

\section{Introduction}{\label{int}}
Low-mass main-sequence stars of spectral type M and L, that is, M and L dwarfs, are abundant in the solar neighbourhood and across the Milky Way. At their low-mass limit they, together with high-mass brown dwarfs, form a group of objects collectively referred to as ultracool dwarfs (UCDs). In 2010, we initiated an astrometric search for exoplanets around 20 nearby UCDs, a project nicknamed `Planets Around L-dwarfs wiTh Astrometry` (PALTA, \citealt{Palta1}). Since then, exoplanet searches around UCDs have been conducted with various observing techniques and have delivered fascinating results, for example, the compact TRAPPIST-1 system hosting several terrestrial planets \citep[host-star spectral type M8\,V,][]{2016Natur.533..221G} {and the system of temperate Earth-mass planets around Teegarden's star \citep[host-star spectral type M7\,V,][]{2019A&A...627A..49Z}},  and the insight that they can indeed host massive giant planets \citep[e.g.][]{2018AJ....156..208J,2019Sci...365.1441M,2020AJ....160...97C}.

Here we report the astrometric discovery of the sub-Jupiter-mass exoplanet candidate \paltab orbiting an early-L dwarf in a $\sim$9 yr orbit, which is a rare example of a giant planet around a UCD and one of the first astrometric exoplanet discoveries with ground-based observations. Data acquisition and analysis are presented in Sect.\ \ref{obs}. The system's orbit and component masses are determined in Sect.\ \ref{orbit}, where we also summarise the overall results of the PALTA survey.
     
\section{Analysis}{\label{obs}}{\label{red}}

\subsection{FORS2 imaging data reduction}\label{FORS2}
The UCD \href{https://simbad.u-strasbg.fr/simbad/sim-basic?Ident=DENIS-P+J125310.8-570924&submit=SIMBAD+search}{DENIS-P J1253108$-$570924} (hereafter \dw or \palta) has a spectral type of L0.5 \citep{Phan} and was observed as part of our long-term astrometric programme on the Very Large Telescope (VLT), aimed at the detection of exoplanets. Observations were obtained from 2011 to 2019 with the FOcal Reducer and low-dispersion Spectrograph (FORS) 2 camera \citep{FORS} in imaging mode with the $I$-band filter. We obtained 28 observational epochs with a total of about 1100 individual exposures. 

The charge-coupled device (CCD) images obtained were calibrated and reduced as described by \citet{Lazorenko2009, PALTA2}. The profiles of the stellar point spread functions (PSF), which have a rather complicated structure,  were approximated  with a sum of three elliptical components, using for the next reduction the most informative component centred at $x_i$, $y_i$ and modelled by 2D Gaussians with standard deviation parameters $\sigma_x$, $\sigma_y$, and correlation $\rho$.  The positions  $x_i$, $y_i$ related to the $i$-th exposure represent the photocentre used for astrometric reduction \citep{PALTA2}, while the other three auxiliary parameters  provide useful information on whether the source is a partially resolved visual binary  {and are the key parameters of the {astrometry of unresolved images (AUI)} analysis} \citep{2025}. Because the PSF profiles of the CCD images are represented by simple 2D elliptic models, they are easily processed in various coordinate system transformations  necessary for the image calibration and, finally, for extracting information on the possible binary geometry of stellar objects (Sect.\,\ref{single}). 

\subsection{Astrometric reduction}{\label{astro}}
The astrometric reduction of the observations was performed with methods proposed by  \citet{LazLaz} that ensure precision of about 0.1--0.2~mas per epoch, primarily by mitigating the atmospheric image motion. These methods have previously been applied successfully to the study of several ultracool binary systems \citep{JS2013, Sahlmann2015,  Sahlmann_palta3,  likely, 2018LUH}. A detailed description of the method was presented in \citet{PALTA2}. 

 The astrometric reduction is performed using approximate CCD positions  $x_{0}$, $y_{0}$ of a dense grid of  reference stars  and assuming that they form a flat, non-distorted reference coordinate system. The coordinate axes are assumed to be oriented along the right ascension  (RA) and declination (Dec) axes. The reference areas are circular and centred on the target. The measured displacements $x_i-x_0$ and  $y_i-y_0$ of photocentre positions  of  each  star within the reference field are modelled by functions related to  two independent sub-spaces: of the CCD $x$, $y$ coordinates, and of time $t$. These displacements are fit by the model 
\begin{equation}
\label{eq:E1}
\begin{array}{@{}l@{}}
 F_{\alpha}(x_i,y_i) + \psi_{\alpha} (t)+ \Delta_{\alpha}= x_i -x_0\\
 F_{\delta}(x_i,y_i) + \psi_{\delta} (t)+ \Delta_{\delta}   = y_i-y_0. 
 \end{array}
\end{equation}
Here, the functions $F(x_i,y_i)$ model the sum of geometric distortion and atmospheric image motion in each frame measured relative to the reference frame, while the change in positions with time $t$ due to proper motion, parallax, chromatic effects, and zero points of positions are included in the function $\psi(t)$. The terms $\Delta_{\alpha}$, $\Delta_{\delta}$ are the residuals of the fit.

The functions $F(x_i,y_i)$ are  full bivariate polynomial  functions of  maximum power  $n$ and include $ (n+1)(n+2)/2$  free fit coefficients formed on the positions of reference stars in the reference epoch. The value of $n$ is defined as  $n=k/2-1$ using astrometric reduction mode $k$, which is an arbitrary even integer.  The primary purpose of the reduction is to eliminate the most important low-frequency components of the image motion spectrum that are removed until the first active  $k$ mode of phase distortion \citep{Lazorenko2009}. We applied reduction with six modes $k$, from six to 16, which corresponds to $n$ in the range of two to seven. In order to ensure the best balance between reference frame noise and  atmospheric image motion, the circular diameters of the reference frames were adjusted to increase  from 70 to 194 arcsec  with $k$ increasing from six to 16.

 The model Eq.\,(\ref{eq:E1}) is solved for each $k$ separately with a least-squares fit to all measurements of  reference stars in all exposures and using weights that depend on the star brightness. Therefore, the fit residuals $\Delta$ of the solution for reference stars are orthogonal both to the basic bivariate  functions in $x_0$, $y_0$ space and to the time-dependent functions. The partial results (proper motion, parallax, fit residuals) obtained from individual $k$ modes are averaged to yield the final results.

The model takes into account some minor effects, such as the relative displacement of the detector chips as described in \citet{PALTA2}. Also, using the absolute positions of \gaia DR1 \citep{DR1}  and later refined with \gaia DR2 \citep{Gaia-Collaboration:2018ae}, we derived and then took into account  the difference in the pixel scale along the FORS CCD $x$, $y$ axes and the inclination of the lower chip of the detector relative to the upper chip \citep{gaia}. 

\subsection{Preliminary fit residuals}
The  parallax derived relative to the reference frame is $\varpi_{\rm rel}=59.83 \pm 0.05$~mas. Because the average parallax of reference stars, as determined from the \gdr{3} \texttt{gaia\_source} catalogue, is $0.30 \pm 0.03$~ mas, the absolute parallax used later for the orbit derivation in  Sect.\,\ref{red} is $\varpi_{\rm abs}=60.13 \pm 0.05$~mas. Our relative and absolute values of parallax and proper motions, and the comparison with \gdr{3}, are shown in Table\,\ref{mu_tbl2}.

The proper motions and parallax determined with FORS2 images  (Table\,\ref{mu_tbl2}) include three components: the quantities measured  relative to the reference stars, the average offsets relative to the absolute materialised by \gaia reference stars (determined for each $k$ reduction mode separately and then merged according to their weights), and the spurious motion $\mu^*$  induced by field geometry distortion that should be subtracted to remove this effect. In Table\,\ref{mu_tbl2}, the column `FORS2 final' contains the parameter values reduced to the absolute system.

\begin{figure}[t]
\resizebox{\hsize}{!}{\includegraphics* [width=\linewidth]{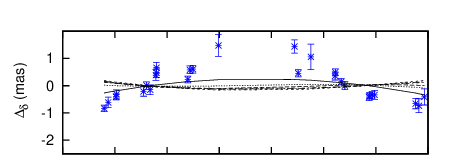}}
\resizebox{\hsize}{!}{\includegraphics* [width=\linewidth]{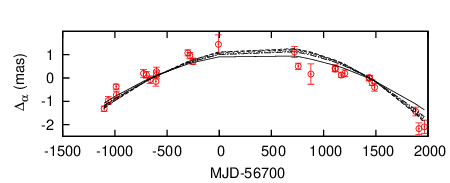}}
\caption{Positional residuals $\Delta_{\alpha}$ in RA (open circles, bottom) and $\Delta_{\delta}$ in Dec (asterisks, top), before removing the additional field distortion effects. The field distortion along the target track over the CCD is shown for different reduction modes as indicated by different line types.}
\label{ocINI}
\end{figure}

\begin{table*}[t]
\caption[]{FORS2 and \gdr{3} astrometry of \palta.}
\centering
\begin{tabular}{@{}ll@{ \, }c@{ \, }c@{ \, }c@{ \, }c@{ \, } c@{}}
\hline
\hline
Parameter   & Unit   & Value (relative) & Offset to absolute &    Correction $\mu^*$            &  FORS2  &  \gdr{3}   \rule{0pt}{11pt}    \\                  
  \hline
$\mu_{\alpha}\cos {\delta}$ &mas yr$^{-1}$&$  -1554.673\pm 0.012$  &$ -6.999\pm 0.046 $  & $ +2.19 \pm 0.30$         & $ -1563.86 \pm 0.30$ & $ -1563.48 \pm 0.14$   \rule{0pt}{11pt}  \\
$\mu_{\delta}$& mas yr$^{-1}$  &$  -430.423 \pm 0.011$  &$ -0.371\pm  0.071$  & $ +0.29 \pm 0.24$         & $  -431.09 \pm 0.25$ & $  -431.53 \pm 0.21$ \\
$\varpi $ & mas & $    +59.832 \pm 0.047$  &$  +0.295\pm  0.026$  & $ ...          $         & $   +60.127  \pm 0.054$ & $    +60.01 \pm 0.18$   \\
\hline
\end{tabular}                                                           
\label{mu_tbl2}                                                                                              
\end{table*}                                                                                                
The positional residuals are shown in Fig.\,\ref{ocINI} and were derived with a typical precision of about 0.2~mas. In both axes they show significant deviations from the standard astrometric model of parallax and proper motions. However, those deviations cannot be directly interpreted as photocentre motions of the source. Instead, additional instrumental effects need to be mitigated first, as described in the following sections.

\subsection{Mitigating position-dependent CCD field distortion}{\label{field}}
\palta has a very high proper motion of 1.5\,arcsec yr$^{-1}$ in RA and because the telescope pointing was kept stable relative to reference stars, this target described a very long track across the CCD with an extent of $\sim$103~px in RA and $\sim$28~px in Dec.  Small time-dependent changes in the CCD field distortion along such a long track can cause biases in the extracted source positions. To mitigate this undesirable effect, the static distortion of the FORS2 reference frame was therefore investigated and taken into account as follows. 

\begin{figure}[t]
\resizebox{\hsize}{!}{\includegraphics* [width=\linewidth]{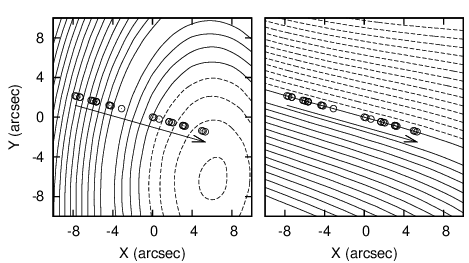}}
\caption{Contour lines of the functions $\Psi_{\alpha}$ (left) and $\Psi_{\delta}$ (right) drawn with 1\,mas and 2\,mas increments, respectively, for the reduction mode $k=12$. The $X$, $Y$ coordinates are centred at the \palta position on the CCD at the modified Julian date (MJD)=56700. The observation epoch positions are shown with open circles and the arrow indicates the motion direction.}
\label{Fg2}
\end{figure}

As an astrometric field calibrator we used the \gdr{3} stellar positions \citep{eDR3cat} and their tangent-plane projections $x_G$, $y_G$ with which we formed the  differences 
\begin{equation}
\label{eq:E2}
\begin{array}{@{}l@{}}
 x_0 - x_G = \Psi_{\alpha} (x_0,y_0) \\
 y_0 - y_G = \Psi_{\delta} (x_0,y_0)  
 \end{array}
\end{equation}
 between the FORS2 reference frame and \gdr{3}. This is the system of conditional equations that we used to model the field distortion of FORS2 with polynomial functions $\Psi$. We derived six  solutions of Eq.\,(\ref{eq:E2}), each related to the currently adopted reduction mode $k$, using the function $\Psi$ formed with the same set of basic functions as adopted for functions $\psi$ in  Eq.\,(\ref{eq:E1}) and with the same optimal field sizes.  The number of basic functions  varied from six ($k=6$) to 36 ($k=16$). {The functions $\Psi_{\alpha}$, $\Psi_{\delta}$ for the mode $k=12$ are shown in Fig.\,\ref{Fg2} for the CCD area  where the target was  measured, for each of the 28 observation epochs.  The shapes of these functions for RA and Dec are very different.}
 
\begin{figure}[t]
\resizebox{\hsize}{!}{\includegraphics* [width=\linewidth]{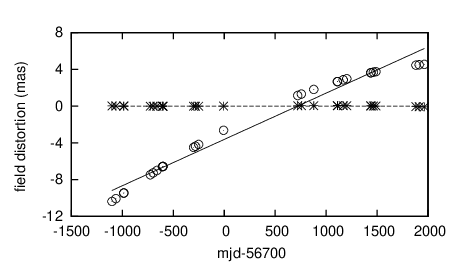}}
\caption{Field distortion $\Psi$  measured along the target's track over the CCD, in  RA (open circles), in Dec  (asterisks) for the reduction mode $k=12$, and their linear fit approximations.}
\label{k12}
\end{figure}

When the field distortion $\Psi$ changes linearly in time along the target's position on the CCD, {its effect} is indistinguishable from proper motion, hence is absorbed by the derived proper motion of the target and does not affect the positional residuals $\Delta$ and causes no problem. 
But a problem arises when $\Psi$  changes  non-linearly with time, which is {illustrated} in Fig.\,\ref{k12} in RA, where the field distortion was measured at the CCD location of \palta for each of the 28 epochs shown in Fig.\,\ref{Fg2}. It contains a component that changes  linearly in time and therefore can be interpreted as a correction $ \mu^*$ to proper motion (Table \ref{mu_tbl2}).  However, non-linear deviations $\Psi(t) -  \mu^* t $ enter the positional residuals and induce a signal that could be mistaken as orbital motion caused by an orbiting companion.  

Figure\,\ref{k12}  displays the distortion derived for the most stable reduction mode $k=12$. For other modes, the deviations $\Psi(t) -  \mu^* t $ are very similar. 
They are quite large in RA and very small in Dec because of a nearly linear change in time of {$\Psi_{\delta}(t)$ (Fig.\,\ref{Fg2}) and because of the target's motion  along a line of `equal distortion`.}

In Fig.\,\ref{k12} 
we show that the effect of the non-linear and variable field distortion is nearly equal to the change of the observed positional residuals in RA. After the subtraction of the CCD field distortion effect, the apparent signal in RA is insignificant. However, the residuals in Dec remain essentially unaffected with an amplitude of about 1~mas. 

{Because the pattern of the astrometric residuals in RA and Dec (Fig.\,\ref{ocINI}) appears similar to the effect of field distortion $\Psi_{\alpha}$ (Fig.\,\ref{k12}), we performed additional investigations into possible causes, for example, the subtle effect of a rotation term or similar, without finding any. 
The corrected residual amplitudes are given in Table\,\ref{OC_tbl} (cf.\ Fig.\ \ref{insky}). We interpret the presence of significant residuals as the orbital signature of a second component of the system.}

\begin{table}[t]
\caption [] {Corrected photocentre positional residuals $\Delta_{\alpha}$, $\Delta_{\delta}$ for \palta.}
{\small
\centering
\begin{tabular}{@{}c@{ \, }c@{ \, } c@{\,  }|@{ \, }c@{ \, }c@{ \, } c@{}}
\hline
\hline
  MJD   &  $\Delta_{\alpha}$ &  $\Delta_{\delta}$  &  MJD   &  $\Delta_{\alpha}$ & $\Delta_{\delta}$ \rule{0pt}{11pt}    \\
        &    (mas)           &    (mas)            &        &    (mas)           &   (mas)     \\
  \hline
  55597.4 &  $-$0.10$\pm  0.11$  & $-$0.92$\pm  0.12$   &57420.4 &  $-$0.04$\pm 0.24$  &  1.48$\pm  0.24$  \rule{0pt}{11pt}  \\
  55636.3 &   0.14$\pm  0.18$  & $-$0.69$\pm  0.19$   &57456.3 &  $-$0.63$\pm 0.13$  &  0.49$\pm  0.13$ \\
  55712.0 &   0.51$\pm  0.11$  & $-$0.46$\pm  0.11$   &57578.0 &  $-$0.85$\pm 0.44$  &  1.09$\pm  0.47$ \\
  55715.0 &   0.15$\pm  0.15$  & $-$0.37$\pm  0.17$   &57809.3 &  $-$0.28$\pm 0.15$  &  0.37$\pm  0.15$ \\
  55974.4 &   0.38$\pm  0.17$  & $-$0.21$\pm  0.18$   &57811.3 &  $-$0.31$\pm 0.12$  &  0.50$\pm  0.12$ \\
  56002.2 &   0.26$\pm  0.14$  &  0.01$\pm  0.14$   &57868.2 &  $-$0.47$\pm 0.10$  &  0.15$\pm  0.12$ \\
  56038.1 &  $-$0.04$\pm  0.17$  & $-$0.13$\pm  0.18$   &57903.0 &  $-$0.33$\pm 0.14$  &  0.00$\pm  0.15$ \\
  56093.0 &  $-$0.23$\pm  0.20$  &  0.41$\pm  0.20$   &58136.3 &   0.01$\pm 0.11$  & $-$0.43$\pm  0.12$ \\
  56098.0 &   0.18$\pm  0.19$  &  0.68$\pm  0.20$   &58140.3 &  $-$0.01$\pm 0.11$  & $-$0.39$\pm  0.12$ \\
  56099.0 &   0.04$\pm  0.17$  &  0.49$\pm  0.18$   &58164.3 &  $-$0.14$\pm 0.12$  & $-$0.35$\pm  0.12$ \\
  56398.2 &   0.41$\pm  0.13$  &  0.28$\pm  0.13$   &58189.3 &  $-$0.27$\pm 0.16$  & $-$0.36$\pm  0.17$ \\
  56418.1 &   0.26$\pm  0.14$  &  0.64$\pm  0.14$   &58581.2 &  $-$0.04$\pm 0.19$  & $-$0.69$\pm  0.19$ \\
  56448.0 &  $-$0.03$\pm  0.14$  &  0.66$\pm  0.14$   &58613.1 &  $-$0.65$\pm 0.25$  & $-$0.77$\pm  0.26$ \\
  56692.4 &   0.40$\pm  0.40$  &  1.54$\pm  0.40$   &58664.0 &  $-$0.37$\pm 0.29$  & $-$0.45$\pm  0.30$ \\
\hline
\end{tabular}                                                                        }
\label{OC_tbl}                          
\end{table}      

\subsection{Verification of the brown dwarf binary system configuration}\label{single}
{Taking into consideration the astrometric evidence (Sect.\,\ref{red}) for the non-single nature of \palta, we considered  the possible configurations of the system that depend on the relative brightnesses $I_B$ and $I_A$  of the secondary (B) and primary (A) components, respectively. Those can be either an ultracool dwarf orbited by an exoplanet with $I_B \ll I_A$, or a pair of ultracool dwarfs with  $0<I_B < I_A$.} Both configurations may result in the same photocentre orbit solution in Sect.\,\ref{orbit}, but distinguishing between them is critically important for the mass estimation of the components. We used the AUI technique \citep{2025} which allowed us to resolve the tight visual binary \dww with a separation of 0.15 arcsec, much below the full width at half maximum (FWHM) of our FORS2 CCD images. {We adapted this method to determine a detection limit for the presence of a second component in the PSF.}

\subsubsection{Basic model of the PSF for binary images}{\label{u}}
The AUI method used to resolve the unresolved  images uses the 2D Gaussian parameters $\sigma_x$, $\sigma_y$, and $\rho$ of CCD images derived as auxiliary terms  during the measurement of the photocentre positions. Those are converted to ellipse parameters  with major $\lambda$ and minor $\sigma$ semi-axes that represent the elliptic PSF shape of the images using the expressions from \citet{ellipse, toolbox}
\begin{equation}
\label{eq:lambd}
\begin{array}{l}
\lambda^2=\frac {\sigma_x^2 + \sigma_y^2}{2} + \sqrt{ \left( \frac {\sigma_x^2 - \sigma_y^2}{2}\right)^2 +  \rho^2 \sigma_x^2 \sigma_y^2}, \\
\sigma^2=\frac {\sigma_x^2 + \sigma_y^2}{2} - \sqrt{ \left( \frac {\sigma_x^2 - \sigma_y^2}{2}\right)^2+\rho^2 \sigma_x^2 \sigma_y^2},\\
\tan{\theta} = \frac{\lambda^2 - \sigma_x^2} { \rho \sigma_x \sigma_y}.\\
\end{array}
\end{equation}

The position angle $\theta$ relative to the $X$-axis of the CCD corresponds to the binary system's orientation. These ellipse parameters completely characterise the profile of the PSF, in particular,  its elongation 
\begin{equation}
\label{eq:uu} 
u =\lambda/\sigma.
\end{equation}
The $\sigma$ parameter depends on the FWHM seeing parameter via the expression   
\begin{equation}
\label{eq:FWHM} 
\sigma = \mathrm{FWHM} /3.1
\end{equation}
that is valid for the FORS2 camera \citep{Lazorenko2009}.

{The key assumption of the AUI method is that the measured PSF of field stars, despite having a rather complicated profile, can be approximated to the photon-noise level by ellipses elongated by optical distortions, atmospheric turbulence, and other effects. Due to this simple geometry of the PSF profile and by applying a sequence of coordinate system rotations and specific calibrations \citep{2025},  the real PSF of any single star can be mapped to an idealised circular PSF. The calibration aims at removing optical deformations and obtaining round images with  $\lambda = \sigma$,  which can be achieved when averaging over a sample of reference field stars. For a single source located at the centre of the reference star group, we expect  $\lambda \approx \sigma$ and $u \approx 1$ to be as expected for a circular image.}

For binary objects, the presence of the secondary light source B at an angular  separation $R$ from the primary A distorts the combined image shape, leading to image elongation. Therefore, differences in the calibrated idealised PSF shapes between the target object and the reference field stars can reveal a faint companion. For $R \leq \sigma$ there is a dependence   
\begin{equation}
\label{eq:uR} 
u = 1+  c \sigma^{-2} R^2+\upsilon
\end{equation}
between $u$ and the angular  separation $R$ between the components of the binary \citep{2025}. Here,  the quantity $\upsilon$ is a noise component, and  the $c$ term  is a function of the luminosity ratio
\begin{equation}
\label{eq:z} 
z = I_B /I_A
\end{equation}
of the A and B components. For equal luminosity components we have $c=0.125$ and $c$ is decreasing towards zero for fainter and fainter companions.  For intermediate cases with $0.4<z<0.8$, {the  approximate expression $c = 0.1194 - 0.123(1-z)^4$ is valid.}

The AUI method  therefore consists of the conversion of the measured  Gaussian parameters $\sigma_x$, $\sigma_y$, and $\rho$ into the parameters $\lambda$, $\sigma$, and $\theta$ of the ellipse. This is followed by a calibration aimed at reducing the systematic distortion of the PSF, the computation of the elongations $u$ for the target object and, finally, fitting these data  with Eq.\,(\ref{eq:uR}) to obtain an estimate of the angular separation $R$. In this procedure, the value of $z$ must either be known independently or estimated otherwise. {The AUI method is capable of resolving very tight companions in CCD images, and here we will use it to reject the binary brown dwarf hypothesis for \dw.}

\subsubsection{Noise in the PSF images}{\label{noise}}
Any instantaneous PSF is  slightly distorted  due to unaccounted systematic optical distortions and fast anisotropic variations of the turbulent atmosphere, which both vary over the field and therefore are not fully removed by calibration. In addition, the knowledge of distortion is limited by measurement errors. These effects introduce  random components $\Delta \sigma_x$, $\Delta \sigma_y$, and $\Delta \rho$  into the 2D Gaussian parameters of the target PSF whose expected true values (free of these random components) are $\sigma_{x,0}$, $\sigma_{y,0}$, and $\rho_0$. 

{After calibration, we expect  to obtain a circular idealised PSF for  a single object with $\sigma_{x,0} = \sigma_{y,0}$, $\rho_0=0$, and thus $u=1$.   The measured   parameters  $\sigma_{x} =\sigma_{x,0} +\Delta \sigma_x$, $\sigma_{y}=\sigma_{y,0}+\Delta \sigma_y$, and $\rho =\rho_{0}+\Delta \rho$, naturally, deviate randomly from the expected values. Even a truly circular PSF } would be  deformed by instantaneous perturbations and becomes an ellipse  with  $\lambda > \sigma$ and thus $u>1$.   For this reason,  the measured $u$ values  are noisy, which  adds some uncertainty in the interpretation of the measurements and in the classification of the object as a binary  or a single star.

While for a strong signal $u \gg 1$ (large $R$) the residuals  $\upsilon$ in Eq.\,(\ref{eq:uR}) are distributed approximately normally, with a standard deviation $\varepsilon$ around its expectation value of $1+  c \sigma^{-2} R^2$, for  weak elongations the distribution of  $\upsilon$  is significantly different.  Because random negative excursions are forbidden due to the implicit requirement of $u>1$, in this case the noise has a one-sided distribution with negative deviations mirrored to positive values. Therefore, it can be expressed as 
 \begin{equation}
\label{eq:R0} 
\upsilon = u_0 + \upsilon' ,
\end{equation}
where  $u_0$ is the (positive) expectation value of a one-sided distribution of $\upsilon $  and $ \upsilon'$ are the remnants of this representation. Also, we can assume that $u_0 \approx \varepsilon$. For single stars with actually zero $u-1$ signal,  the model Eq.\,(\ref{eq:uR}) for the measured elongations is modified to
 \begin{equation}
\label{eq:R00} 
u =1+ u_0 + \upsilon'.
\end{equation}
In intermediate cases, with non-zero but weak $u-1$ signal when the noise dominates,  it is difficult to differentiate between the PSF corresponding to a single star or to a binary.
 
To mitigate  the ambiguity of the solution, we applied  Monte Carlo simulations of the terms $u$ and $\theta$, and {took the opportunity to use a comparison field star of comparable brightness located close to the target. The calibration of  this star's PSF was performed  exactly as for the target, with, however, a slightly different set of reference stars. The benefit of using the comparison star is that it does not require assumptions on the random errors of the measurements, which are difficult to model adequately. We only assumed that these errors are statistically indistinguishable between \palta and the comparison star, allowing a small systematic difference due to the small offset on the CCD. }

The simulated distribution of $u$ and $\theta$  depends on the magnitude of the random components $\Delta \sigma_x$, $\Delta \sigma_y$, and $\Delta \rho$. We modelled it, adopting as a reasonable guess the values  used by \citet{2025} when modelling the  $u$ and $\theta$  distributions for the unresolved binary system \dww observed with FORS2 imaging.   Thus, $\Delta\sigma_x$ and   $\Delta\sigma_y$ were set to be random values with  a standard deviation that changes between 0.019~px   and 0.07~px  with  seeing  changing from the best to the worst conditions;  for $\Delta\rho$, we set  a standard deviation of 0.006 irrespective of seeing. 

\subsubsection{Measured elongations  of FORS2 images}
\begin{figure}[t]
\resizebox{\hsize}{!}{\includegraphics* [width=\linewidth]{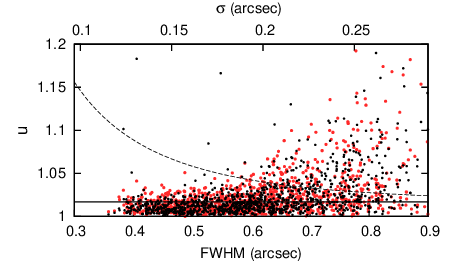}}
\caption{Elongations $u$ as a function of seeing for  \palta ({red}) and for the comparison star (black). The solid line shows the noise level $u_0$ and  the   expected dependence for the alternative binary configuration (Sect.\,\ref{orbit}) is shown with a dashed line.}
\label{U_L2}
\end{figure}

{\label{teta}}
{After calibrating the measured PSF, using five field stars of comparable brightness within 200~pixels of \palta,  we computed the elongations $u$  for the science object and for the comparison star. Figure\,\ref{U_L2} shows the distributions  of these elongations $u$ as a function of $\sigma$ and  FWHM. These distributions are very similar for the field star and for \palta.} Any signal, if present, would be weak, thus the noise of $u$ values is better represented by  Eq.\,(\ref{eq:R0}), with which we find $u_0 \approx 0.015 $  for both the comparison star and \palta for seeing FWHM<0.7 arcsec.

We detected a small systematic excess $\Delta u_0$ when comparing the $u_0$ value of \palta with that of the comparison star.  For seeing between 0.4 and 0.6 arcsec, this excess was found to vary between $0.001$ and $0.002$ with an uncertainty of $\sim50$\,\%. We did not find any dependency of $\Delta u_0$ on $\sigma$ as  would be expected from Eq.\,(\ref{eq:uR}) if it were caused by the binary signature. Therefore, we attributed this excess to the limitations of  calibrating sources at different CCD locations.

If we assume that \palta is a binary system of two UCDs of comparable brightness, then, according to  Eq.\,(\ref{eq:uR}), the measured values of  $u$  should contain a component that decreases with $\sigma$ like the function shown by a dashed line in Fig.\,\ref{U_L2}. {This line corresponds to  the  signal induced in $u$ that is compatible with the astrometric orbit parameters derived in Sect.\,\ref{orbit} in the scenario where \dw is an ultracool binary. It is clear that this dashed line does not match our observations, which allows us to reject the ultracool binary scenario with high confidence.}

By fitting  the $u$ values  measured in $\sim 700$ images with FWHM$ <0.7$ arcsec  using   Eq.\,(\ref{eq:uR}) and the representation  Eq.\,(\ref{eq:R0}) for  $\upsilon $, we found that  for a binary system with equal-luminosity components, or $c=0.125$, the separation $R$ cannot exceed $R_{\rm max} =0.023$ arcsec at 3-sigma confidence. This is in agreement with  \citet{2025}, who found that the smallest detectable separation $R_{\rm max}$ resolved  with a few  hundred FORS2 images  is about  0.03 arcsec.  In Fig.\,\ref{U_L2}, the corresponding signal {is $u-1-u_0=0.004$ } at 0.4 arcsec seeing. 

{The dependence of the elongation $u$ on seeing leads to the conclusion  that \palta is either a very tight binary with $R<0.023$ arcsec (a scenario rejected in Sect.\,\ref{masses}) or a UCD with an exoplanet companion. In both cases, its PSF is indistinguishable from that of a single comparison star. Because this conclusion is of critical importance for this study, we confirm it in the following section.}

\subsubsection{Distributions of the PSF parameters}{\label{u_teta}}
The noise in the images affects the shape of the calibrated PSF, inevitably transforming any circular image into an ellipse. When due to noise,  $\Delta\sigma_x> \Delta\sigma_y$  is registered in the measured PSF
of a single object with $\sigma_{x,0} = \sigma_{y,0}$, then the result is an elongation of the PSF profile along the $X$ axis, which corresponds to a position angle of $\theta=0$ of the semi-major axis. In the opposite case, the  ellipse is elongated along the $Y$ axis, corresponding to $\theta= 90$\degr.  Thus, the 2D distribution of the parameters $u$ and $\theta$ for a single star is represented by two peaks, one near $\theta =0$ and one at $\theta =90$\degr. At a 50\% occurrence rate of $\Delta\sigma_x> \Delta\sigma_y$, the number of events in each of these peaks is equal. Fluctuations in the correlation parameter ${\rho}$ will smear these peaks out. However, the distribution of errors in the PSF parameters can deviate from the normal distribution. As a result, the occurrence rate of $\Delta\sigma_x> \Delta\sigma_y$ may be above or below 50\%, thus the number of events in the peaks are expected to be slightly unequal.

\begin{figure}[t]
\resizebox{\hsize}{!}{\includegraphics* [width=\linewidth]{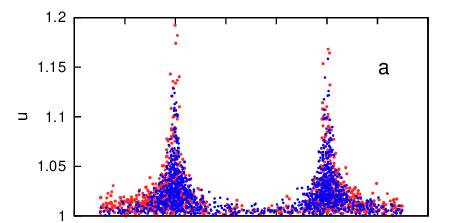}}
\resizebox{\hsize}{!}{\includegraphics* [width=\linewidth]{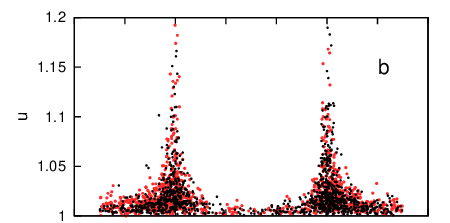}}
\resizebox{\hsize}{!}{\includegraphics* [width=\linewidth]{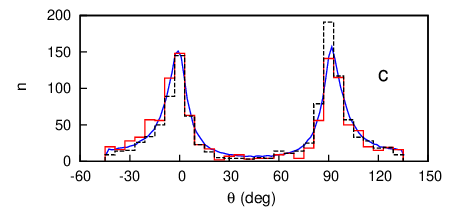}}
\caption{Observed 2D distribution of the parameters $u$ and $\theta$ for a) \palta (red circles) and a simulated set of PSFs (blue circles); b) \palta (red circles) and a nearby field star (black dots); c) histogram of $\theta$  for \palta (red solid line), the comparison star (black dashed line), and the simulated distribution (blue solid curve).}
\label{UTrnd}
\end{figure}

 Figure\,\ref{UTrnd} displays the result  for a simulated single object with a circular PSF, which has two peaks near  0\degr and 90\degr, as expected. The 2D distributions are shown for both our target and simulated data. The agreement between both distributions is good and is also seen in  Fig.\,\ref{UTrnd},b where we compare these distributions for \palta and for the comparison star. That means that \palta is either a single object or a binary with an undetectable faint companion, thus very small $R$ or small $z$.  

\begin{figure}[t]
\resizebox{\hsize}{!}{\includegraphics* [width=\linewidth]{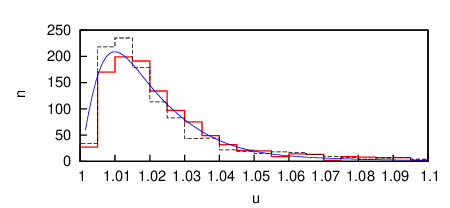}}
\caption{Distribution of the number $n$ of elongations $u$  for  \palta (red steps),   the comparison star (dashed black steps), and  the simulated distribution (blue solid line).}
\label{Ugist}
\end{figure}

The distributions in Fig.\ \ref{UTrnd},a,b are shown for  illustrative purposes. To quantify the statistical agreement between these distributions, we computed the histograms of the position angles $\theta$  (Fig.\,\ref{UTrnd},c) and elongations $u$ (Fig.\,\ref{Ugist}) for \palta,  the comparison star, and the simulated set of PSFs, and then applied an $\chi^2$ test. For the comparison star, we noticed a moderate asymmetry in the distribution, with a $\sim$20\% excess in the  $45 <\theta < 135$\degr range compared to the  $-45< \theta < 45$\degr range  (Fig.\,\ref{UTrnd},b). As commented above, for a comparison star with a circular PSF, this feature can be related to an excess of cases with $\Delta \sigma_y > \Delta \sigma_x$ caused by an imperfect calibration based on only five reference field stars and deviations of the PSF measurement error distribution from the normal law. For \palta a much smaller excess of $\sim$5\% is found  around $\theta=0$\degr as a consequence of the larger number of  $\Delta \sigma_x > \Delta \sigma_y$ cases for this source. The difference in excess is due to slightly different CCD locations of these sources, which implies different PSF distortions. Considering that the distribution of $\theta$ angles for \palta is more symmetric than for the field star (Fig.\,\ref{UTrnd},c), we find that the former is in good agreement with the circular PSF expected for a single object.

In addition to comparing the $\theta $ distributions, we analysed the difference in the width of the  $\theta$ distributions near the peaks. To robustly determine this metric, we preliminarily removed the asymmetry effect discussed above and caused by the different event numbers of $\Delta \sigma_x $ and $ \Delta \sigma_y$, by merging the right ($\theta > 45$\degr) to the left  ($\theta < 45$\degr)  area of distributions, thus forming a distribution limited to $\theta$ angles between $-$45 and +45\degr, which is robust to the unequal event numbers $\Delta \sigma_x $ and $ \Delta \sigma_y$. After this correction, we applied the $\chi^2$ test  to quantify the difference in distribution width for \palta and the field star. Using Poisson statistics for the number of measurements in histogram bins, the inconsistency of the distributions is  expressed  by  $\chi^2$ computed from differences of counts in these distributions, using only bins with sufficient data points.  This resulted in $\chi^2=17.2$ with 15 degrees of freedom (DOF), well within the 20\% tolerance limit, which does not allow us to reject the hypothesis that these distributions are identical within statistical fluctuations.

{The distributions of $\theta $ for \palta and the comparison star are seen to be slightly asymmetric relative to their peaks at 0\degr and 90\degr, but with the same asymmetry pattern for both objects, which, as noted above, can be due to imperfect calibration. To model this feature in the simulated set of PSFs, we  added a constant $-$0.002 offset to the parameter $\rho$, which produced a good match to observations (Fig.\,\ref{UTrnd},c). Thus we  demonstrated that the histograms for \palta and the comparison star  can be reproduced with appropriate simulations for an actually circular PSF.}

We applied an $\chi^2$ test to find possible differences in the distribution of elongation ($u$) for \palta and the field star. The corresponding  histograms  for the full set of images  shown in Fig.\,\ref{Ugist}  produced  a large value of $\chi^2=27.9$ with 14 DOF, which was found to be caused by images obtained in bad seeing.  By restricting the measurements to good seeing (FWHM $\leq 0.6$ arcsec, satisfied by roughly half of the images), the $\chi^2$ value was reduced to 16.6 with 13 DOF. These good-seeing images are the ones most sensitive to the image elongation effect. When computing $\chi^2$, we also discarded images with $u<1.01$ that are rare (cf.\ Fig.\,\ref{Ugist}) and likely related to calibration issues rather than actual image elongation.

{Figure \ref{Ugist} also shows the simulated distribution for a synthetic binary with zero separation ($R=0$), which illustrates how the  Gaussian noise in the measurements enters the statistics of the measured elongations $u$. }The peak of the distribution  is close to $u_0\approx 0.015$, and the wings correspond to the distribution of  the residual noise $ \upsilon'$ introduced in Eq.\,(\ref{eq:R0}).

All analyses presented in this section indicate that \palta is not spatially resolved with our data because its and the  field star's histograms  are identical within statistical fluctuations. The photocentre astrometry, however, indicates the presence of a companion that therefore could be either an object of negligible brightness such as an exoplanet or a binary companion of approximately equal brightness ($z \approx 1$) but at a spatial separation too small {to affect the PSF shape significantly. In Sect.\,\ref{orbit} we use an orbital fit to distinguish between these scenarios and to reject the latter case}.

\section{Results and discussion}{\label{orbit}}
\subsection{Orbital solution and exoplanet mass}
\subsubsection{Parameters of the orbit}{\label{orbpar}}
The orbit of \palta was derived by fitting the measured photocentre deviations from the linear single-source model $\Delta_{\alpha}$, $ \Delta_{\delta}$ {given in Table\,\ref{OC_tbl}}  with the model
\begin{equation}
\label{eq:orbF}
\begin{array}{@{}rcr@{}}
BX + GY +x_0 +\Delta\mu_{\alpha}T & =   & \Delta_{\alpha}\\
AX + FY +y_0 +\Delta\mu_{\delta}T & =   & \Delta_{\delta}, \\
\end{array}
\end{equation}
where $X$, $Y$ are the elliptical orbital coordinates of the photocentre.  These equations include the Thiele-Innes parameters  $A$, $B$, $F$, $G$,  the coordinate offsets $x_0$, $y_0$, and the proper motion corrections $\Delta\mu_{\alpha}$, $\Delta\mu_{\delta}$, which reduce the proper motion of the photocentre, within the observation period,  to that for the barycenter motion  relative to reference stars. The Thiele-Innes parameters refer to the photocentre orbit with the semi-major axis $a_\mathrm{ph}$ and which, for a very faint companion, coincides with the barycentre orbit of the primary. 

This system of non-linear equations was solved  following  the approach of \citet{Lucy}, by splitting the orbital parameters into a group of non-linear parameters, which include the eccentricity $e$, orbital period $P$, and time of periastron passage $T_0$, and a group of linear parameters $A$, $B$, $F$, $G$. The solution was derived by exploring the non-linear parameters over a wide range of values,  in this way turning Eq.\,(\ref{eq:orbF}) into a linear system that can be solved  with a standard least-squares procedure. The resulting values $A$, $B$, $F$, $G$ were then transformed into the Campbell orbital elements. 

\begin{figure}[t]
\resizebox{\hsize}{!}{\includegraphics* [width=\linewidth]{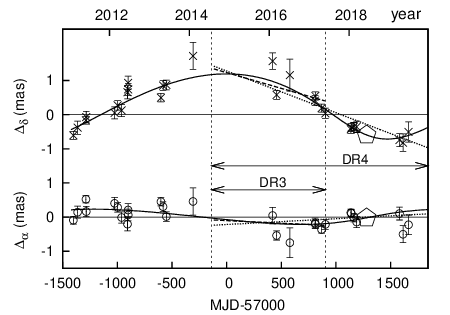}}
\caption{Measured positional residuals in right ascension (circles, bottom) and in declination (crosses, top). The photocentre-to-barycentre orbital model motion (solid curves) and the periastron position (pentagon) are shown. Linear fits to the orbit segments corresponding to \gdr{3} (dashed) and \gdr{4} (dotted lines) are shown and the corresponding observational periods are marked by arrows.}
\label{insky}
\end{figure}

\begin{figure}[tbh]
\centering
\includegraphics* [width=1\linewidth,clip=true,trim={-3mm -3mm 0mm 0mm}]{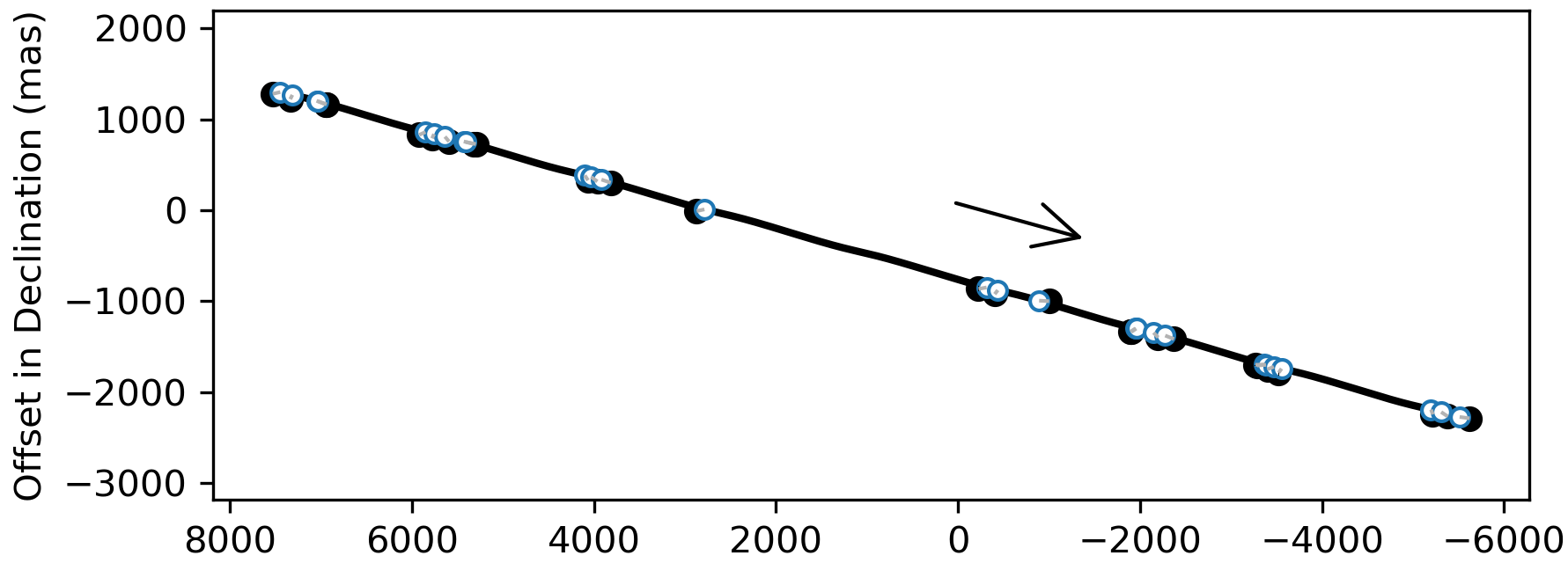}
\includegraphics*[width=\linewidth]{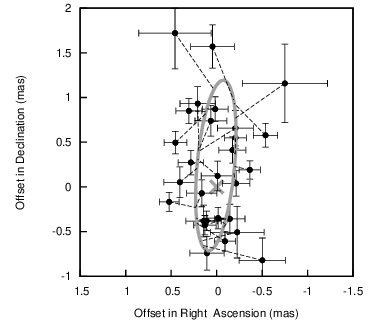}
\caption{\emph{Top}: Parallax and proper motion of \palta in the plane of the sky, showing the predominant eastwards motion of $>$12 arcsec over the observation timespan. \emph{Bottom}: Orbital motion (grey curve) of the \palta photocentre around the system's barycentre (grey cross) in the sky after subtracting the
single-source model. The grey curve corresponds to the orbital parameters. The dashed lines connect the observed and computed locations.}
\label{fig:orbit}
\end{figure}

The best-fit model for the photocentre orbital motion relative to the barycentre, expressed by the terms  $BX + GY$ and $AX + FY$ in right ascension and declination, respectively, is shown in Fig.\,\ref{insky} as a function of time {and projected on the sky in Fig. \ref{fig:orbit}}. The observed photocentre shifts relative to the barycentre $\Delta_{\alpha}- x_0 -\Delta\mu_{\alpha}T$ and $\Delta_{\delta} -  y_0 -\Delta\mu_{\delta}T$ are plotted.  

The set of parameters that produce the best fit were chosen as the final solution and are given in Table\,\ref{tbl}  with  $1\sigma$  confidence intervals estimated as  described by \citet{Lucy}. Because of the sometimes large uncertainties, we also list the lower and upper confidence intervals.  The timespan of the FORS2 data is 8.4 years, and the best-fit period is slightly longer at 8.8 years. The observations do not cover a full orbital period or a clear inflection point; hence, the best-fit period corresponds to the shortest compatible period and the true period is likely longer, as indicated by the large upper confidence interval on this parameter. This is a known feature of fitting astrometric orbits when the orbital period is not covered.
      
\begin{table}[t]
\caption[]{Orbital parameters and component mass estimates of the \palta system.} 
{ 
{\centering
\begin{tabular}{@{}l@{ \,}ccc@{}}
\hline
\hline
Parameter                                &             Value                        & Lower margin      &   Upper margin        \\
\hline                                                                                                                     
$e$                           &  $  0.29_{-0.21}^{+ 0.36}$               &  0.075   & 0.65    \rule{0pt}{11pt} \\
$T_P$ (d)                   &  $  58280_{-900}^{+240}             $    &    57376   &   58524   \rule{0pt}{11pt} \\
$P$ (yr)                      &  $     8.8_{-1.1}^{+8.4}         $       &      7.7   &    17.2    \rule{0pt}{11pt} \\
$a_\mathrm{phot}$ (mas)       &  $     0.97_{-0.13}^{+0.83}   $          &     0.84   &    1.80   \rule{0pt}{11pt} \\
$i$ $({\degr})$                  &  $     102.7_{-5.7}^{+7.5}   $           &     97.0   &   110.2   \rule{0pt}{11pt}  \\
$\omega$ $({\degr})$             &  $    329_{-106}^{+34}   $               &      222   &     362   \rule{0pt}{11pt}  \\
$\Omega$ $({\degr})$             &  $     174.4_{-5.7}^{+8.0}   $           &    168.6   &   182.4    \rule{0pt}{11pt}  \\
$\Delta \mu_{\alpha}$ (mas yr$^{-1}$)        &  $     -0.017_{-0.030}^{+0.093}$         &   $-$0.047   &   0.077   \rule{0pt}{11pt}  \\
$\Delta \mu_{\delta}$ (mas yr$^{-1}$)        &  $      0.044_{-0.080}^{+0.26}$         &   $-$0.036   &    0.30     \rule{0pt}{11pt}  \\
\hline 
      \multicolumn{4}{c}{ Solution with $M_A =90\,M_\mathrm{Jup}$ } \rule{0pt}{11pt} \\
$a_\mathrm{rel}$ (mas)       &  $    114_{-16}^{+97}    $               &      98    &   211   \rule{0pt}{11pt}  \\
$a_\mathrm{rel}$ (au)        &  $      1.89_{-0.26}^{+1.62} $           &    1.63    &   3.51   \rule{0pt}{11pt}  \\
$M_B$ $(M_{\rm Jup})$           &  $    0.77_{-0.070}^{+0.20}   $           &       0.70 &   0.97    \rule{0pt}{11pt}  \\
\hline 
      \multicolumn{4}{c}{ Solution with $M_A =52\,M_\mathrm{Jup}$ } \rule{0pt}{11pt} \\
$a_\mathrm{rel}$ (mas)     &  $     95_{-13}^{+81}    $            &       82   &     176   \rule{0pt}{11pt}  \\
$a_\mathrm{rel}$ (au)      &  $      1.58_{-0.22}^{+1.35} $        &    1.36    &   2.93    \rule{0pt}{11pt}  \\
$M_B$ $(M_{\rm Jup})$          &  $    0.54_{-0.048}^{+0.14}   $        &       0.49 &   0.67   \rule{0pt}{11pt}  \\
\hline                                       \end{tabular}                                }
\label{tbl}                                  
}                                            
\end{table}           

Another consequence is that the parameter uncertainties are quite large, for example, the eccentricity is almost unconstrained with values between 0.075 and 0.65 at 1$\sigma$. The orbital period estimate lies between 7.7 and 17.2 years. However, this does not prevent us from obtaining reliable estimates of the mass $M_B$ of the companion as shown in Sect.\,\ref{masses}.

To derive these confidence intervals, we modified the original method presented by \citet{Lucy} which takes into account the correlation between a pair of parameters ($B$, $G$) and between ($A$,  $F$) only, thus the covariance matrix $\vec{ D}$ of a linear subset of size $4\times 4$.  We modified this method because for the particular case of  Eq.(\ref{eq:orbF}), the  linear  ($A$, $B$, $F$, $G$) group of parameters is expanded and also includes  $x_0$, $y_0$,  $\Delta\mu_{\alpha}$, and $\Delta\mu_{\delta}$;  therefore,  the matrix $\vec{ D}$ is of size $8 \times 8$. This is similar to \citet{2018LUH}, where we solved a similar problem for the matrix  $\vec{ D}$ of dimension $7\times 7$ using the Jacobi transformation \citep{Press}  of this  matrix that leads  to its diagonalized form  $\vec {U = V^{T} D V}$ where $\vec {V}$ is a matrix whose columns are eigenvectors of $\vec{ D}$. The values of $\sqrt{U}$ formed from the diagonal elements   of $\vec{U}$  are the eigenvalues of $\vec{D}$.  For this transformation,  the orthogonal variables $\hat z$  \citep[Sect.\,A.4]{Lucy} are transformed to the displacement of the linear subset of parameters \citep[Eq.(A.11), Eq.(A.15)]{Lucy} using the expression $\vec {V  U^{1/2} \hat z }$ where $\vec{U^{1/2}}$ is the diagonal matrix of the elements $\sqrt{U}$. 

\subsubsection{Estimation of component masses}{\label{masses}}
A VLT/Ultraviolet and Visual Echelle Spectrograph (UVES) \citep{Dekker:2000aa} spectroscopic observation of \palta was obtained on 18 April 2013 with an exposure time of 2800 s and a resolving power of 34540. An S/N of 9.3 was estimated by the automatic pipeline for the red part of the spectrum (565.5 -- 946.4 nm). Visual inspection of the spectrum found no significant absorption due to the Li I resonance doublet and a weak absorption of H-alpha. A radial velocity of $-33.4 \pm 6.0$ km s$^{-1}$ was estimated from the position of the H-alpha line.

\palta has a spectral type of L0.5 and based on the absence of Li absorption in our UVES spectrum we estimated the mass $M_A$ of the primary {to be confined between 52\,$M_{\rm Jup}$  and  90\,$M_{\rm Jup}$, where we used estimates from \citet{Martin2022} and the dynamical mass for LSPM J1735+2634B \citep[their Table\,11]{2017ApJS} , respectively, including uncertainties}. Under the assumption that the secondary is dark ($I_B=0$), its mass $M_B$ was derived using  the orbital period $P$,  the absolute parallax $\varpi_{\rm abs}=60.13 \pm 0.05$~mas (Sect.\,\ref{red}),  the photocentre semi-major axis $a_\mathrm{ph}$,  Kepler's law $M= (a_\mathrm{\rm rel} /\varpi_{\rm abs})^3/ P^2$ for the total mass $M=M_A+M_B$,  and the relation  $ a_\mathrm{rel}  = a_\mathrm{phot} f^{-1}$  between the  photocentre and the semi-major axis of the relative orbit  $a_\mathrm{rel}$, where $f=M_B/M$ is the mass function. The mass $M_B$, which meets all above relations, was found as the solution of the equation
\begin{equation}
\label{eq:MB}
(a_\mathrm{phot}/\varpi_{\rm abs})f^{-1} = (M_A+M_B)^{1/3}P^{2/3}.
\end{equation}

Table\,\ref{tbl} presents the secondary mass derived  for both the lower and upper primary mass limit. Despite the poor precision of the orbital parameters, the companion mass is constrained with an uncertainty of  10--20\% (cf.\ comments by \citealt{Lucy2014}) for a fixed primary mass $M_A$. Depending on primary mass, the best-fit companion mass $M_B$ ranges between 0.54\,$M_{\rm Jup}$ and 0.77\,$M_{\rm Jup}$, that is, firmly in the exoplanet mass range. To account for this range of possibilities, we adopted an estimate of $M_B = {0.65} \pm 0.12$  $M_\mathrm{Jup}$ for the mass of the companion to \palta. The mass ratio $q=M_B/M_A$ in all these cases is about  $q_0= 0.009$ -- 0.010.  

If we consider the alternative to a dark companion, that is, a configuration of two UCDs and $I_B>0$, then $ a_\mathrm{rel}  = a_\mathrm{phot} (f-\beta)^{-1}$ and  Eq.~(\ref{eq:MB}) is modified to  
\begin{equation}
\label{eq:MB2}
(a_\mathrm{phot}/\varpi_{\rm abs})(f-\beta)^{-1} = (M_A+M_B)^{1/3}P^{2/3},
\end{equation}
 which contains the fractional flux $\beta=I_B/(I_A+I_B)$.   It has two solutions. The first one is found with $I_B=0$, or $\beta=0$, and corresponds to the exoplanet case already analysed above.   To restrict the range of possible parameter values in the second {scenario} with $\beta>0$, we used Eqs.\ (\ref{eq:MB}) and (\ref{eq:MB2}) to come to an approximate expression $f-\beta \approx q_0(1+q)^{-1/3}$ where $q_0 $ is the mass ratio for the exoplanet configuration. It follows that the quantity $f-\beta $, which defines the ratio of the photocentre $ a_\mathrm{phot}$ to the relative $a_\mathrm{rel} $ orbit, is  as small as $q_0$ or about 0.01. Hence, the system components should have a comparable mass ($q \approx 1$) and luminosity  ($I_B/ I_A \approx 1$). {To derive numerical solutions, we solved Eq.(\ref{eq:MB2}) with all possible values of $f$ and $\beta$, using the approximate relation $z=I_B/I_A = (M_B/M_A)^{\gamma}$ between the masses and the luminosities. For L dwarfs, the value of $\gamma$ is known to depend strongly on the age of the system. Using the brown dwarf evolutionary models of \citet{chabrier23}, we found that for ages below 1~Gyr, the mass-luminosity relation can be approximated by a $\gamma$ value between two and four (cf.\ Figure\,10 of \citealt{2025}). For older UCDs with ages up to 10~Gyr, the dependence is much steeper, corresponding to a larger value of  $\gamma$ \citep[their Figure\,3]{chabrier23}. The uncertainty in the $\gamma$ value is not problematic here because  a solution of Eq.\,(\ref{eq:MB2}) exists at $q \approx 1$ and is therefore independent of the actual value of $\gamma$. For $\gamma =2.0--4.0$, which implies ages below 1~Gyr, the derived solutions are all within a narrow range of $q \simeq 0.97--0.99$ and  $f-\beta \simeq 0.007--0.008$, depending  on the adopted primary mass.  These solutions derived for the best estimates of $a_\mathrm{phot} $ and $P$  (Table\,\ref{tbl}) lead to a large relative orbit size $a_\mathrm{rel}  = a_\mathrm{phot} (f-\beta)^{-1}$ of about 2~au and separations of $ R\sim 0.12--0.14$ arcsec, which are forbidden by the constraint of  $ R< 0.023$ arcsec, determined in Sect.\,\ref{single} based on the largest separations compatible with a single PSF. For ages greater than 1~Gyr, we derived even larger estimates of $R$. Therefore, the hypothetical binary configuration of \dw can be excluded.}

\begin{figure}[t]
\resizebox{\hsize}{!}{\includegraphics* [width=\linewidth]{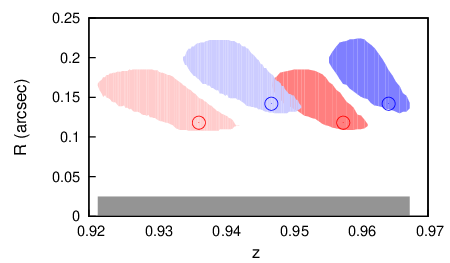}}
\caption{Angular relative separation $R$ as a function of flux ratio $z$ for simulated binary systems that are compatible with our astrometric orbit,  derived for $M_A={90}\,M_{\rm Jup}$ (blue),  $M_A=52\,M_{\rm Jup}$ (red), and for $\gamma=4.0$ (darker shading) and  $\gamma=2.0$ (lighter shading). The corresponding results for the best-fit astrometric orbit are shown with circles. The allowed space of solutions that is consistent with the measured elongations of the FORS2 images (Sect.\ \ref{teta}) is shown as a dark-shaded rectangle. }
\label{areas}
\end{figure}
 
{We verified that these conclusions hold given the relatively large uncertainty of the astrometric orbit (Table\,\ref{tbl}) and searched for a parameter space of the binary scenario near or below the threshold set by the AUI analysis in Sect.\,\ref{single}. We used the simulated set of Eq.\ (\ref{eq:orbF}) solutions values of   $a_\mathrm{phot}$ and $P$ drawn within the  $\pm 1 \sigma$ confidence intervals for the astrometric  solution  and generated using the technique described by \citet{Lucy}. For every sample in this space, we computed the solution assuming the binary scenario. The results in terms of angular separation $R$ and flux ratio $z=I_B/I_A$ are shown in Fig.\ \ref{areas} for the two extreme values of primary mass $M_A$ and two values of $\gamma$. Almost independently of the adopted values $M_A$ and $\gamma$, the smallest angular separations $R$ lie significantly above the maximum value allowed by the AUI analysis. This finding could be confirmed with future high-resolution imaging of this system.}

We were awarded VLT Nasmyth Adaptive Optics System Near-Infrared Imager and Spectrograph (NaCo) time as part of ESO programme {0103.C-0428} to observe \dw with the main goal of determining whether the source can be spatially resolved as a UCD binary, as is the case for \dww \citep{2025}. Unfortunately, these observations could not be acquired before the NaCo instrument was decommissioned.

{Finally, from Eq.\ (\ref{eq:uR}) we found that the elongations $u$ corresponding to the smallest separation $R$ in Fig.\,\ref{areas} should be about 1.07--1.09 at seeing with FWHM$= 0.4$ arcsec.  This is  much larger than the observed range shown in Fig.\,\ref{U_L2}, and such strong elongations would have been detected even within a single FORS2 exposure. Together, these considerations reject the scenario where \palta is a binary UCD and therefore support the scenario in which the companion is a sub-Jupiter mass exoplanet that we designate \paltab.}

\subsubsection{Orbit detectability with \gaia}{\label{gaiacomp}}
\palta was {observed by} \gaia and the orbital motion discovered with FORS2 could in principle be detected with \gaia astrometry as deviations from the standard single-source model. \palta is catalogued in \gdr{3} \citep{GaiaCollaboration:2016aa,2023A&A...674A...1G} as source {Gaia DR3 6060561079334177664}, with a mean magnitude of $G \approx 18.2$\,mag. It has a standard six-parameter astrometric solution \citep{Lindegren:2021vn} whose proper motions and parallax are in broad agreement with our initial determination by \citet{Palta1}. 

For \gdr{3}, \palta was observed by \gaia between 25 July 2014 and 28 May 2017. Assuming that our FORS2 proper motions are unbiased, a linear fit limited to the FORS2 orbit segment during that timespan should correspond to the  difference between the \gaia and FORS2 proper motions. This is because the orbital period is about three times longer than the \gdr{3} timespan. Following this computation (cf.\ Fig.\,\ref{insky}), we expect that the \gaia proper motions are more negative by $0.172$~mas yr$^{-1}$ in right ascension and by $0.293$~mas yr$^{-1}$ in declination. This corresponds well, at least qualitatively, to the parameters given in Table \ref{mu_tbl2} but given the associated uncertainties, it should be considered marginal. Other possible non-single star indicators published in \gdr{3}, for example,  RUWE=1.097 as an indicator for unresolved binarity or the 	
ipd\_gof\_harmonic\_amplitude=0.0095 as an indicator of partially resolved binarity, are inconspicuous and suggest that this source appears as being single in \gdr{3}. 

To confirm this, we simulated the \gaia astrometric time series by using the \palta sampling and scan angles returned by the \gaia observation forecast tool ({\tt GOST}\footnote{\url{https://gaia.esac.esa.int/gost/index.jsp}}) and injecting the astrometric motion corresponding to our best-fit orbital model, that is, the proper motions and parallax from Table \ref{mu_tbl2} and the orbital parameters from Table \ref{tbl}, using the \texttt{pystrometry} package \citep{johannes_sahlmann_2019_3515526}. We assumed an astrometric uncertainty of a single CCD measurement of 1 mas, which roughly corresponds to the expectation for a $G=18.2$\,mag source in \gdr{3} \citep{2023A&A...674A..10H} and also for \gdr{4} \citep{brown_2025_17158744}, and injected white noise of the same amplitude into the simulated time series.

We then fitted these simulated time series with the standard 5-parameter model of positions, parallax, and proper motions and searched for significant deviations from this model by computing the periodogram of this model's residuals using the \texttt{kepmodel} package \citep{2022A&A...667A.172D}. This method has been shown to be very sensitive to non-linear signals such as drifts and curvature due to long-period orbital motion in the time series \citep{2024ApJ...963...60P}. When using the simulated \gdr{3} time series with a timespan of $\sim$3 years, there is no indication of any non-linear signal from the periodogram analysis. This confirms the earlier hypothesis that the data timespan is too short and the measurement uncertainties are too large for \gdr{3} to detect the orbit's non-linear signature. The only effect is the bias in the \gdr{3} proper motions as discussed above.

We repeated this analysis for simulated \gdr{4} time series that cover $\sim$5.5 years and a larger fraction of the orbit (cf.\ Fig.\,\ref{insky}) and find the same result: there is no significant indication of non-linear motion. Finally, we did the same for simulated \gdr{5} time series that cover $\sim$10.5 years with almost double the number of individual observations compared to \gdr{4}. The periodogram of the simulated \gdr{5} data shows a very significant peak at the period corresponding to our orbital period of \paltab, indicating that the planet's signature will be detectable in the mission's final data release. This is the consequence of covering more than one full orbital period with many astrometric measurements that have roughly the same precision as the signal's semi-amplitude.

In summary, we found that the \gdr{3} results are compatible with the properties of the \palta system as determined with FORS2 and that we will have to wait for \gdr{5} to independently confirm the astrometric orbit caused by the exoplanet candidate \paltab.

\subsection{Summary of the results of the PALTA survey}{\label{summ}}
As the PALTA survey is nearing completion, we summarise below the main results that we obtained. Table~\ref{sum} presents an overview of the survey targets and the respective results. 
\subsubsection{High-precision astrometry}
The PALTA survey was made possible by the advances in high-precision astrometry that can be achieved with seeing-limited imaging observations with very large telescopes \citep{LazLaz,Lazorenko2009,PALTA2}. For the first time, astrometric accuracies of 0.1 mas were achieved for a survey programme \citep{Palta1}. This exceptional performance allowed us to determine extremely precise parallaxes of ultracool dwarfs \citep{Palta1}, probe the astrometric precision of \gdr{1} \citep{gaia}, characterise the mechanical stability of the CCD mosaic of the FORS2 instrument camera \citep{2019A&A...629A.113L}, and predict a performance limit of 5 $\mu$as for seeing-limited observations with extremely large telescopes \citep{2024A&A...689A..94L}.

First demonstrated for the VLT, similar high-precision astrometry was successfully demonstrated on other 10-m class telescopes such as the Gran Telescopio Canarias  \citep[GTC,][]{Sahlmann:2016aa} and Gemini-North and Gemini-South \citep{2020MNRAS.495.1136S}.
{The astrometric parameters, that is, proper motions and parallaxes, determined for all the survey targets are presented in Table~\ref{sum}. These parameters are determined relative to reference stars and thus are not absolute. The absolute parameters of \dw/\palta are given in Table~\ref{mu_tbl2}.}

\begin{table*}[t]
\caption {PALTA survey targets' astrometric parameters and detection limits for additional companions.}
{
\centering
\begin{tabular}{@{}l@{ \, }cccc@{ \, }l@{ \, }l@{ \, }c@{ \, }c@{  \,  }l@{}}
\hline
\hline
DENIS-P &$\mu_{\alpha}\cos\delta$ &$\mu_{\delta}$   &$\varpi $  & $\Delta T  $\tablefootmark{a} & Astr.   & AUI     & $R_\mathrm{max} $ & $R_\mathrm{max}$ &  Object   \rule{0pt}{11pt}         \\
        &(mas yr$^{-1}$) &(mas yr$^{-1}$) & (mas)&            (yr)                           & signal  & signal  & (arcsec)                  &  (au)               &   type\tablefootmark{b}    \\
\hline
J0615493-010041  &  199.01 $\pm$  0.11 & -56.13  $\pm$ 0.12 & 45.25$\pm$ 0.11   &   1.2 & No &    No           &       0.034        &      0.75    & s  \rule{0pt}{11pt}\\
J0630014-184014  &  326.46 $\pm$  0.03 &-502.54  $\pm$ 0.03 & 50.58$\pm$ 0.07   &   3.3 & Yes&    Yes          &       ...          &      ...     & b   \\
J0644143-284141  &  223.03 $\pm$  0.12 & -80.39  $\pm$ 0.11 & 24.76$\pm$ 0.11   &   1.3 & No &    No           &       0.030        &      1.21    & s   \\
J0652197-253450  & -235.70 $\pm$  0.11 &  87.26  $\pm$ 0.09 & 61.50$\pm$ 0.08   &   1.3 & No &    No           &       0.027        &      0.44    & s   \\
J0716478-063037  &  -11.01 $\pm$  0.14 & 148.33  $\pm$ 0.12 & 40.61$\pm$ 0.13   &   1.3 & No &    No           &       0.027        &      0.66    & s   \\
J0751164-253043  & -877.09 $\pm$  0.01 & 144.20  $\pm$ 0.01 & 56.08$\pm$ 0.06   &   6.4 & No &    No           &       0.020        &      0.36    & s   \\
J0805110-315811  & -237.54 $\pm$  0.09 &  87.31  $\pm$ 0.08 & 42.13$\pm$ 0.10   &   1.3 & No &    No           &       0.030        &      0.71    & s   \\
J0812316-244442  &  136.67 $\pm$  0.03 &-142.80  $\pm$ 0.02 & 47.28$\pm$ 0.11   &   5.9 & No &    No           &       0.027        &      0.57    & s   \\
J0823031-491201  & -154.86 $\pm$  0.07 &   7.96  $\pm$ 0.06 & 48.27$\pm$ 0.11   &   2.3 & Yes&    No           &       0.027        &      0.56    & b   \\
 J0828343-130919 &  -577.07$\pm$  0.12 &   26.34 $\pm$ 0.10&  85.26$\pm$ 0.15   &   1.3 & No &    No           &       0.027        &      0.32    & s   \\
 J1048278-525418 &  -233.18$\pm$  0.07 &   45.76 $\pm$ 0.06&  36.01$\pm$ 0.09   &   2.4 & No &    No           &       0.036        &      1.00    & s   \\
 J1157480-484442 &   -58.47$\pm$  0.03 &  -18.22 $\pm$ 0.03&  34.64$\pm$ 0.08   &   5.2 & No &    No           &       0.029        &      0.84    & s   \\
 J1159274-524718 & -1052.53$\pm$  0.05 & -129.56 $\pm$ 0.06& 104.76$\pm$ 0.09   &   3.0 & No &    No           &       0.029        &      0.28    & s   \\
 J1253108-570924 & -1554.67$\pm$  0.01 & -430.42 $\pm$ 0.01&  59.83$\pm$ 0.05   &   8.4 & Yes&    No           &       0.023        &      0.38    & p   \\
 J1520022-442242 &  -619.01$\pm$  0.02 & -378.99 $\pm$ 0.02&  54.00$\pm$ 0.09   &   5.1 & No &    No           &       0.038        &      0.70    & wb   \\
 J1705474-544151 &   -82.15$\pm$  0.04 &   28.25 $\pm$ 0.05&  37.68$\pm$ 0.06   &   2.3 & No &    No           &       0.028        &      0.74    & s   \\
 J1733423-165449 &    73.64$\pm$  0.03 &  -34.46 $\pm$ 0.02&  56.43$\pm$ 0.07   &   6.1 & No &    No           &       0.024        &      0.43    & s   \\
 J1745346-164053 &   110.07$\pm$  0.05 &  -96.72 $\pm$ 0.05&  50.87$\pm$ 0.07   &   2.3 & No &    No           &       0.025        &      0.49    & s   \\
 J1756296-451822 &    45.51$\pm$  0.01 & -176.96 $\pm$ 0.01&  43.62$\pm$ 0.04   &   8.3 & Yes&    Yes          &       ...          &      ...     & b   \\
 J1756561-480509 &    79.67$\pm$  0.04 &   34.91 $\pm$ 0.05&  46.89$\pm$ 0.06   &   2.3 & No &    No           &       0.027        &      0.58    & s   \\
\hline     
\end{tabular} 
\tablefoot{
\tablefoottext{a}{Timespan of observations.}
\tablefoottext{b}{Object type classification: s - single; b - binary; p - with planet; {wb - wide binary (Sect.\,\ref{65})}.}
}
}\label{sum}                                                            
\end{table*}      

\subsubsection{Detection limits for additional companions}
In Table~\ref{sum} the four targets for which we detected astrometric perturbations are indicated. We discuss those systems briefly in the following subsections.
For the remaining targets we applied the AUI method \citep{2025} to search for the subtle signatures of marginally resolved systems.
We performed the same analyses as described in Sect.\ \ref{single} for all targets.

After calibration of the images as described in Sect.\,\ref{u}, we computed the elongations $u$ and fit them with  Eq.\,(\ref{eq:uR}) where  the noise component $\upsilon$ for low signals was  represented by Eq.\,(\ref{eq:R0}). Except for \dwtwo and \dww, for which we detected astrometric signals, the AUI method failed to identify any new binary configurations. The remaining 18 targets, therefore, have a PSF that is compatible with one single light-emitting source and are flagged by 'No' in the `AUI signal` column of Table~\ref{sum}. {The presence of a periodic astrometric signal means that the object is either a tight binary (such as \dwnine) or a system with a planet (such as \dw). The distinction between these cases requires additional investigation like that described in Sect.\,\ref{masses}.}

Using the same technique, we were able to determine an upper separation limit of $R_{\mathrm max}$ for a hypothetical equal-luminosity binary configuration. The component separations $R_{\mathrm max}$ were estimated with 3-sigma confidence as described in Sect.\,\ref{teta} and refer to the equal-luminosity components when the parameter $c$ in Sect.\,\ref{u} is equal to 0.125. In other cases when the companion is fainter, the value of $R_{\mathrm max}$ increases and, for example, for $I_B=0.1 I_A$ it is approximately two times larger.

\subsubsection{New ultracool binary discoveries}
The first published PALTA result was the discovery of the binary brown dwarf \dwnine in a $\sim$0.7 yr orbit \citep{JS2013}. Later, it was shown that the  system must be juvenile with lithium detected in its optical spectrum and a spectral binarity signature in the near-infrared \citep{Sahlmann2015}. 
The AUI analysis for this system reveals a small signal in the $u$ values below the detectability limit for the binary structure. This is consistent with the separation and magnitude difference of the system components estimated in \citet{JS2013,Sahlmann2015}.

The discovery of the multiple nature of \dwtwo was described in \citet{Sahlmann_palta3}, where an astrometric orbit with a period of 3 years was determined. Follow-up studies and spatial resolution of the system with aperture masking revealed that the system is likely young and that the companion itself is a tight binary \citep{likely}.
The elongation $u$ value of the system's PSF is moderately large and about 4.5 times over its uncertainty determination,  confirming  the  non-single structure of this object. PALTA astrometry of \dww revealed a signature compatible with an orbiting exoplanet, but follow-up adaptive optic imaging and independent AUI analysis revealed the binary nature of this system of two juvenile brown dwarfs in a $\sim$27 year orbit \citep{2025}.

 \subsubsection{Extrasolar planets around ultracool dwarfs}
 Initially, the PALTA survey established that super-Jupiter exoplanets are rare around ultracool dwarfs. Specifically, we derived planet exclusion limits that set an upper limit of 9\% on the occurrence of giant planets with masses $>5\,M_\mathrm{Jup}$ in intermediate-separation (0.01–0.8~au) orbits around M8–L2 dwarfs.

\begin{figure}[t]
\includegraphics[width=\linewidth]{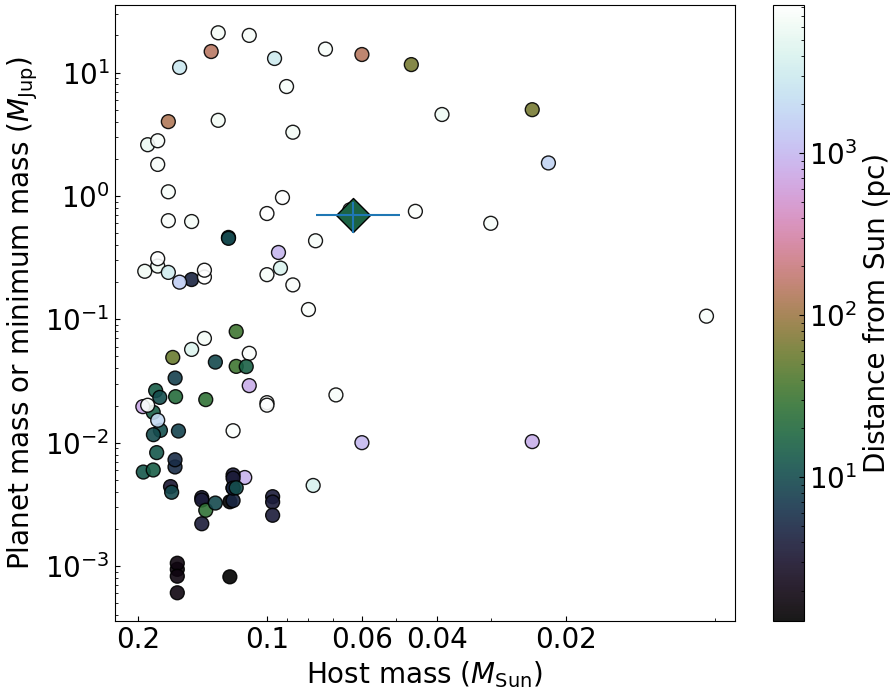}
\caption{Planet mass as a function of host mass for systems included in the NASA Exoplanet Archive \citep{2025PSJ.....6..186C} queried on 4 March 2026. Only systems with a host mass $M<0.2\,M_\sun$  and available system distance are shown. \paltab is indicated with a diamond and the symbol colour indicates the system's distance from the Sun.}
\label{fig:context}
\end{figure}

 In this work, we announce the discovery of \paltab, a sub-Jupiter
exoplanet candidate with a period of $\gtrsim$9 years around the L0 dwarf \dw at a distance of $\sim$17 pc. Figure \ref{fig:context} shows \paltab in the context of other exoplanets around very low-mass stars and ultracool dwarfs. In the shown host mass -- exoplanet mass space, \paltab is embedded in a population of giant planets around low-mass hosts that have been discovered predominantly by micro-lensing. The figure illustrates that \paltab is one of the first examples of a Jupiter-mass planet around a nearby ($<$20 pc) potentially substellar host.

 \subsubsection{Ultracool multiplicity occurrence rate}{\label{65}}
Among the 20 ultracool survey targets, PALTA discovered three binaries with separations $<$5~au and one exoplanet in an orbit  $\gtrsim$2~au.
The target sample also included the wide $\sim$20~au binary \dwfifteen \citep{Burgasser:2007gd} but no astrometric perturbation or signature was found in the AUI analysis around either component. 

We therefore obtain an occurrence rate of tight ($<$5~au) ultracool binaries of $15^{+11}_{-5}$ \%. The combined occurrence of tight and wide binaries is $20^{+12}_{-6}$ \%, and the overall multiplicity rate, including exoplanets, is $25^{+12}_{-7}$ \%. {These estimates are in broad agreement with ultracool multiplicity statistics obtained from a variety of techniques \citep[e.g.][]{Bouy:2003kx,Burgasser:2007ix,Blake:2010lr,Winters:2019aa,2019ApJ...883..205B,2023ASPC..534..275O}}

The significant number of tight ultracool binary systems found in our survey indicates that astrometric surveys are very efficient at uncovering this population of binaries. The all-sky \gaia survey is sensitive to those systems as demonstrated in \gdr{3} \citep{b2023AA...674A..34G,2024MNRAS.533.3784B} and the upcoming data releases \gdr{4}\footnote{\url{https://www.cosmos.esa.int/web/gaia/data-release-4}} and \gdr{5} will deliver a uniform and unprecedented census.

\section{Conclusions}{\label{conc}}
We presented the discovery of the exoplanet candidate \paltab in a $\sim$9-year orbit around the L0 dwarf \dw/\palta based on long-term astrometric monitoring with FORS2/VLT. {The astrometric orbit of the photocentre allows for two possible scenarios: a binary system composed of two near-equal brightness ultracool dwarfs or a single ultracool dwarf orbited by a much fainter exoplanet. Using detailed modelling of the observed optical images of \dw, we could reject the former scenario, which leads to the discovery of the exoplanet candidate \paltab.} The estimated planet mass is between 0.54\,$M_{\rm Jup}$ and { 0.77}\,$M_{\rm Jup}$,  depending on the uncertain mass of the ultracool dwarf host. The system mass ratio is about 0.010, and the relative separation is 1.6--3.4~au, which corresponds to 0.08--0.20 arcsec. 

\palta b {is announced as a candidate} primarily because of the uncertainties in the orbital solution due to the incomplete orbit coverage. Independent confirmation may prove challenging because of the faintness of the ultracool dwarf host. The expected radial-velocity amplitude, for example, is $\sim$50~m\,$s^{-1}$. High-contrast imaging could be another avenue towards confirmation, but it would have to be sufficiently performant to detect a sub-Jupiter exoplanet at a spatial separation of 0.08--0.20\,arcsec from an L0 dwarf. We expect that the astrometric orbit of \palta will be determined independently with \gdr{5} data that will not be available  before the end of 2030\footnote{\url{https://www.cosmos.esa.int/web/gaia/release}}. If confirmed, \paltab would be one of the very few known examples of a giant planet orbiting a nearby ultracool dwarf. 

\begin{acknowledgements}
{Based on observations made with ESO telescopes at the La Silla Paranal Observatory under programme IDs  087.C-0567,  089.C-0397,  091.C-0083,  596.C-0075, and 0103.C-0428.}

The authors thank Didier Queloz for his contributions to the PALTA survey and the referee, Adam J.\ Burgasser, for his thorough review and comments that helped improve the presentation of the results.
ELM was funded by a European Research Council grant nicknamed Substellar (project number: 101054354).

This work has made use of data from the European Space Agency (ESA) mission
{\it Gaia} (\url{https://www.cosmos.esa.int/gaia}), processed by the {\it Gaia}
Data Processing and Analysis Consortium (DPAC,
\url{https://www.cosmos.esa.int/web/gaia/dpac/consortium}). Funding for the DPAC
has been provided by national institutions, in particular the institutions
participating in the {\it Gaia} Multilateral Agreement.

This research has made use of the NASA Exoplanet Archive, which is operated by the California Institute of Technology, under contract with the National Aeronautics and Space Administration under the Exoplanet Exploration Program, of the SIMBAD database, operated at CDS, Strasbourg, France \citep{Wenger:2000ve}, of \texttt{astropy}, a community-developed core Python package for Astronomy \citep{Astropy-Collaboration:2013aa}, \texttt{scipy} \citep{Jones:2001aa}, \texttt{numpy} \citep{Oliphant2007}, \texttt{ipython} \citep{Perez2007}, \texttt{pandas} \citep{jeff_reback_2022_6408044}, \texttt{matplotlib} \citep{hunter2007}, and \texttt{astroquery} \citep{2019AJ....157...98G}, and of \texttt{pystrometry}, an open source Python package for astrometry timeseries analysis \citep{johannes_sahlmann_2019_3515526}.
\end{acknowledgements}

\bibliographystyle{aa}
\bibliography{palta1b_v2} 
\end{document}